\documentclass[sigconf]{acmart}

\AtBeginDocument{%
  }

\setcopyright{acmlicensed}
\copyrightyear{2024}
\acmYear{2024}
\setcopyright{rightsretained}
\acmConference[UIST '24]{The 37th Annual ACM Symposium on User Interface Software and Technology}{October 13--16, 2024}{Pittsburgh, PA, USA}
\acmBooktitle{The 37th Annual ACM Symposium on User Interface Software and Technology (UIST '24), October 13--16, 2024, Pittsburgh, PA, USA}
\acmDOI{10.1145/3654777.3676403}
\acmISBN{979-8-4007-0628-8/24/10}

\newcommand*{\add}{\textcolor{black}}

\begin{document}

\title{Silent Impact: Tracking Tennis Shots from the Passive Arm}

\author{Junyong Park}
\email{jyp0802@kaist.ac.kr}
\affiliation{
  \institution{School of Computing, KAIST}
  \city{Daejeon}
  \country{Republic of Korea}}
\orcid{0000-0003-3962-4585}

\author{Saelyne Yang}
\email{saelyne@kaist.ac.kr}
\affiliation{
  \institution{School of Computing, KAIST}
  \city{Daejeon}
  \country{Republic of Korea}}
\orcid{0000-0003-1776-4712}

\author{Sungho Jo}
\email{shjo@kaist.ac.kr}
\affiliation{
  \institution{School of Computing, KAIST}
  \city{Daejeon}
  \country{Republic of Korea}}

\newcommand{\sysname}[0]{Silent Impact}


\begin{teaserfigure}
    \centering
    \includegraphics[width=.92\linewidth]{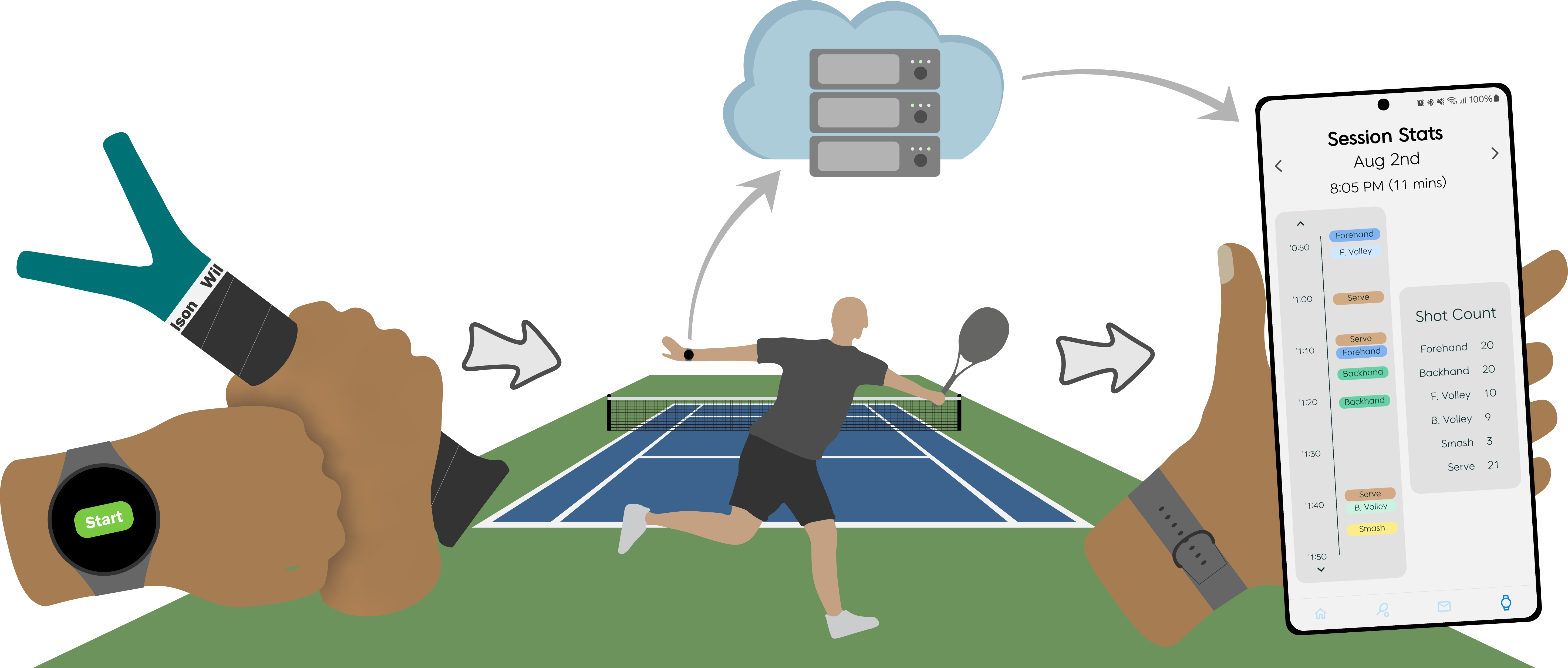}
    \caption{
    Unlike most analytical products that require sensors to be placed on the player's dominant arm, \sysname{} analyzes tennis shots from the passive arm. Users start the app on their smartwatch and proceed to play tennis as usual. After the match, a breakdown of their shots is provided on their smartphone. The rightmost image is a screenshot of our app.
    }
    \label{fig:shot_tracker_flow}
    \Description{}
\end{teaserfigure}

\begin{abstract}

Wearable technology has transformed sports analytics, offering new dimensions in enhancing player experience. Yet, many solutions involve cumbersome setups that inhibit natural motion. In tennis, existing products require sensors on the racket or dominant arm, causing distractions and discomfort. We propose \sysname{}, a novel and user-friendly system that analyzes tennis shots using a sensor placed on the passive arm. Collecting Inertial Measurement Unit sensor data from 20 recreational tennis players, we developed neural networks that exclusively utilize passive arm data to detect and classify six shots, achieving a classification accuracy of 88.2\% and a detection F1 score of 86.0\%, comparable to the dominant arm. These models were then incorporated into an end-to-end prototype, which records passive arm motion through a smartwatch and displays a summary of shots on a mobile app. User study (N=10) showed that participants felt less burdened physically and mentally using \sysname{} on the passive arm. Overall, our research establishes the passive arm as an effective, comfortable alternative for tennis shot analysis, advancing user-friendly sports analytics.

\end{abstract}

\maketitle

\section{Introduction}

Wearable technology has become integral in our lives over the last decade \cite{ferreira2021wearable, wu2019wearable}. Watches, once mere timekeepers, now analyze sleep patterns, track hand wash frequency, calculate calories burned, and monitor various daily activities \cite{chang2018sleepguard, li2018wristwash, bhattacharya2022leveraging}. This trend has also sparked innovation in other 
wearables, like glasses and earphones, to explore new functionalities that enhance our daily experiences \cite{opaschi2020uncovering, verma2021expressear}. The sports industry has also embraced wearable technology, introducing devices that track movements, analyze actions, and provide feedback to the athletes \cite{aroganam2019review}. Such solutions are offered through devices of various forms; some integrate sensors into existing sportswear and gear like socks, goggles, rackets, or gloves, while others require new apparel like vests, straps, or suits.

In tennis, products that analyze swing motion and track tennis shots have emerged, either attached to the racket’s handle or strings \cite{headtennissensor, babolatplay, zepp}, or worn on the wrist like a watch or a wristband \cite{smash}. With the widespread adoption of wrist-worn devices globally \cite{ccs_insight_2022}, the use of commercial smartwatches and fitness trackers has also been explored \cite{taghavi2019tennis, lopez2019site}. However, the common characteristic of these approaches is that the device needs to be placed on the dominant arm’s side. This can result in hindered wrist movement or imbalances in weight, potentially causing discomfort for tennis players who are particularly sensitive to such subtleties \cite{brody2000player, carello1999perceiving, davids2004sensitivity}. Coupled with the possible burden of acquiring a new device, these factors make them less appealing to casual sports enthusiasts, resulting in the discontinuation of many commercialized products.

\begin{figure}[t]
    \centering
    \includegraphics[width=\linewidth]{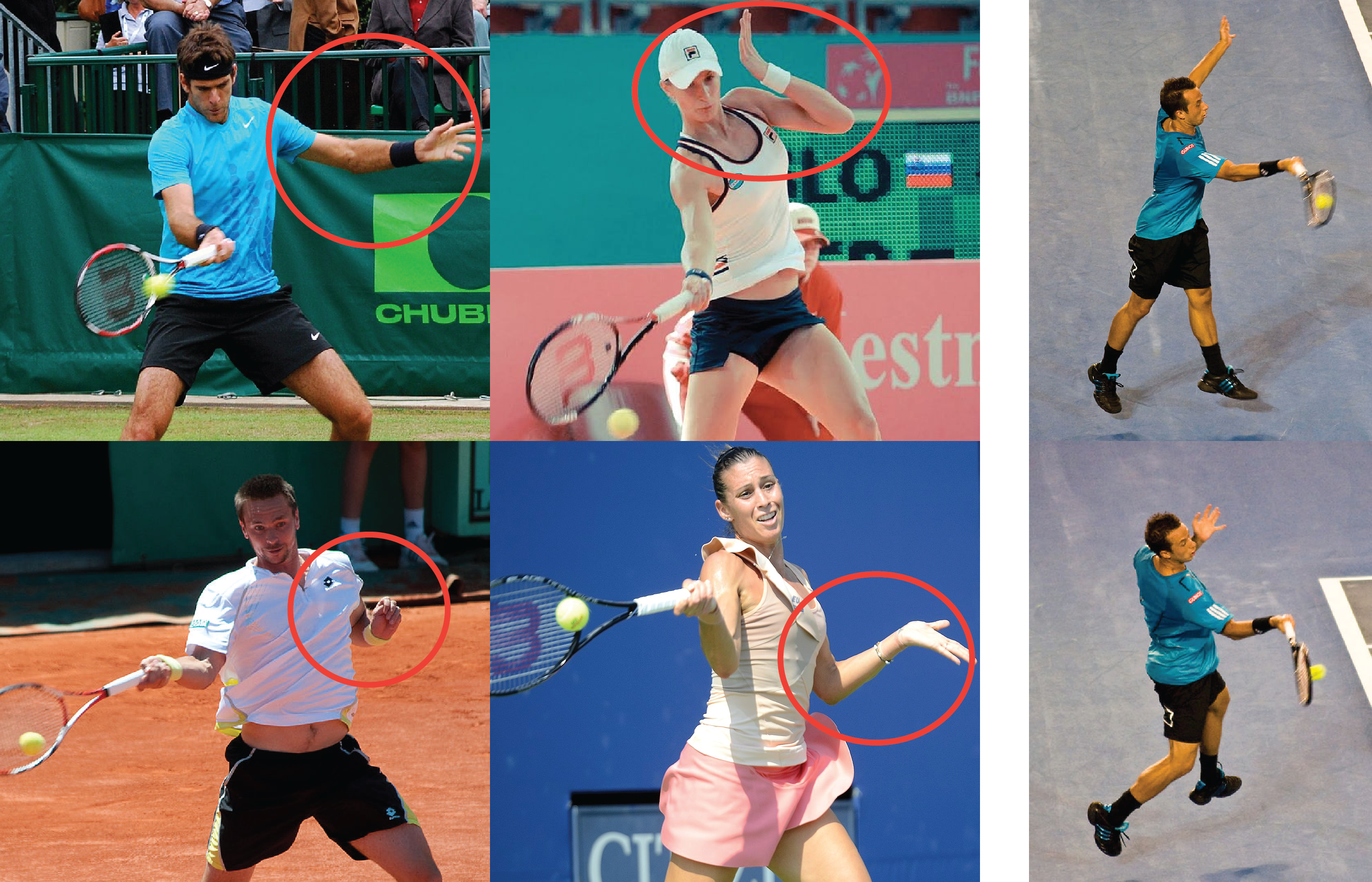}
    \caption{Comparison of the pose of the passive arm (circled in red) when hitting a forehand. Photos are taken roughly at the same moment --- right before the ball impact. Left: Different professional players have different forms for their passive arms. Right: The same player can show varying motion depending on the situation (e.g., ball speed, location). Images are taken from \cite{tennisform1,tennisform2,tennisform3,tennisform4,tennisform5,tennisform6}}
    \label{fig:pro_form}
    \Description{}
\end{figure}

A preliminary survey examining the smartwatch usage patterns of 40 recreational tennis players revealed that all smartwatch owners wear their devices on their passive (non-dominant) arms, with 90\% maintaining this during tennis play. Therefore, utilizing the sensors of a smartwatch worn on the passive arm would provide a more user-friendly solution. However, this presents a complex challenge: when the racket strikes the ball, sensors on the racket or the dominant arm can capture jerks caused by the sudden change in velocity, which cannot be captured from the passive arm. Additionally, the movements of the passive arm can vary significantly, as it doesn't directly influence how the ball is hit (Figure~\ref{fig:pro_form}). \add{Although pose estimation research has demonstrated that body pose can be estimated using Inertial Measurement Unit (IMU) data from various body parts, including the passive arm \cite{devrio2023smartposer, mollyn2023imuposer}, these methods often necessitate multiple devices or are limited to static motions, rendering them impractical for dynamic sports such as tennis.}

To address this challenge, we present \sysname{}, an innovative tennis shot-tracking system that leverages the passive arm data. \sysname{} records IMU data on a smartwatch as a user plays tennis, detects and classifies tennis shots through neural networks, and provides a breakdown of shot types and timeline on a mobile phone (Figure~\ref{fig:shot_tracker_flow}). Specifically, all shot instances within continuous tennis play are first detected (i.e., Shot Detection) which are then classified into one of the six distinct shot types -- \textit{serve}, \textit{smash}, \textit{forehand stroke}, \textit{backhand stroke}, \textit{forehand volley}, and \textit{backhand volley} (i.e., Shot Classification). Utilizing the IMU sensor --- a standard feature in commercial smartwatches --- ensures that our approach can be seamlessly applied to commercial smartwatches, providing users with a convenient and user-friendly experience.

To build \sysname{}, we first gathered IMU sensor readings from 20 recreational tennis players of varying tennis experience and styles. The participants performed six distinct types of tennis shots for the Shot Classification dataset, and ten participants from the cohort played rallies and casual matches for the Shot Detection dataset. We collected data from both the dominant and passive arms, using the dominant arm data to label the moment of each shot and compare the efficacy of each sensor location.

We then employed neural network models for shot detection and classification. Our shot classification model, inspired by established techniques for dominant arm data, featured 1-dimensional convolutional blocks. To adapt to passive arm data, the model used the Fourier transform, decomposing signals into frequency bands for enhanced pattern recognition. For shot detection, we chose the MS-TCN model, known for its effectiveness in action segmentation, particularly in discerning temporal dependencies. Integrating these models, we developed an end-to-end prototype application.\footnote{\add{The models and dataset collected in this work are available at \url{https://github.com/jyp0802/Silent-Impact}}}

In terms of shot classification, we achieved an average 5-fold cross-validation accuracy of 88.2\% using sensor data from the passive arm, which is only 1.9\% lower than that achieved with dominant arm data. Similarly, for shot detection, the passive arm data exhibited an accuracy of 95.6\% and an F1 score of 86.0\%, which are 2.9\% and 8.8\% lower than the dominant arm, respectively. These findings underscore the potential of leveraging the passive arm for efficient and robust tennis shot analysis.

In a user study with 10 participants, participants played tennis with our smartwatch either on their dominant arm or passive arm. Our results show that participants felt significantly less mental and physical burden when using \sysname{} in both the setup process and when playing tennis, compared to when the dominant arm had to be used. They found the system non-intrusive and convenient, while wearing the device on the dominant arm was generally uncomfortable, citing issues such as hand-device contact, added weight, and constrained wrist movement from tight straps. 

Through this study, we demonstrate the technical and practical feasibility of leveraging the motion of the passive arm for tennis shot tracking. This breakthrough simplifies the tracking process for future wearable technology and alleviates the burdens associated with using sensors on the dominant arm. \sysname{} offers a user-friendly solution for tennis enthusiasts seeking to effortlessly monitor their activity, opening up new possibilities for using subsidiary motions as proxies for detailed motion analysis.
\section{Related Work}

In this section, we review both academic research and industry products related to human motion analysis in sports as well as general applications. We explore various methodologies employed in tennis and other sports. Several commercial products discussed here have been discontinued but are included for a comprehensive scope of related work. We also look at initiatives aimed at enhancing the usability of such technology.

\subsection{Motion Analysis for Tennis}
Tennis activity analysis has attracted attention from both academia and industry, resulting in diverse methods categorized into camera-based, racket-attached, and body-worn sensors. While academic studies offer in-depth human motion insights, numerous commercial products have emerged as real-world applications. However, several such devices mentioned below have now been discontinued.

The SwingVision app employs computer vision techniques to provide metrics such as shot type count, longest rally, and shot speed. Unique to this modality, it also tracks where the ball lands on the court \cite{swingvision}. However, such techniques require an elevated camera setup and often struggle with challenges like variable lighting and occlusion. Numerous studies have also examined tennis activity through videos or 3D motion data. FarajiDavar et al. developed a system for identifying tennis shots from broadcast videos, classifying them into three categories \cite{farajidavar2011transductive}. Gourgari et al., on the other hand, created a comprehensive 3D motion database, categorizing a diverse array of tennis shots into 12 distinct types \cite{gourgari2013thetis}. For those looking to capture more nuanced data, some researchers have turned to motion capture systems \cite{skublewska2020learning, skublewska2023temporal}. These systems offer higher levels of accuracy but are less practical for everyday use due to their immobility and complex setup.

To capture more detailed movements, some approaches employ motion sensors positioned nearer to the source of the action, the tennis racket. QLIPP is a commercial device that attaches to racket strings like a dampener \cite{qlipp}. Its close proximity allows it to discern shot types, speed, spin, strike point on the racket, and shot consistency. Yet, due to its rigidity and size, users reported changes in the racket's vibration, weight, and balance \cite{qlippreview}. Other devices mount to the racket handle's base \cite{zepp, sonysmarttennissensor}, which can shift the racket's weight and interfere with those gripping the handle's end. Some technologies are integrated within the racket handle \cite{pei2017embedded, babolatplay, headtennissensor}. Although these typically preserve the racket's balance and weight, they are available in only select models.

Sensors worn on the body offer advantages by eliminating the need to modify rackets. Studies indicate that data from the dominant wrist can facilitate various tennis motion analyses, including shot detection, type classification, and consistency evaluation \cite{wu2022real, kos2018tennis}. Several commercial products, such as Smash wearables \cite{smash}, Babolat POP \cite{babolatplay}, and PIQ \cite{piq}, adopt this approach, designing wristbands with embedded sensors like accelerometers and gyroscopes. These devices can tally shot types, measure racket speed and wrist rotation, and assess impact point consistency. Additionally, research has ventured into using audio signals \cite{ganser2021classification} and mainstream smartwatches \cite{lopez2019site}. Some have also incorporated multiple sensors on different body parts (e.g., chest, wrist, pelvis, shin, knees) \cite{ahmadi2009towards, benages2019detection, yang2017tennismaster, pivot} or combined multiple modalities such as a smart wristband (IMU) and a smartphone (depth sensor) \cite{jia2021swingnet} for more detailed analysis or tennis stroke and service motions. Nonetheless, it remains consistent that any comprehensive tennis shot analysis necessitates a sensor on the dominant wrist.

\subsection{Wearable Sensors in Other Sports}
Sensors are becoming increasingly prevalent in both professional sports and everyday fitness activities. For example, in football and American football, organizations like FIFA and the NFL employ vests equipped with sensors to monitor players' movements, tracking metrics like position and speed during games \cite{catapult}. MLB pitchers use sensor-laden armbands during training to analyze their pitching techniques \cite{pulsethrow}. For everyday fitness enthusiasts, brands such as Fitbit \cite{fitbit} and Garmin \cite{garmin} have popularized wrist-worn devices that capture a broad spectrum of health and fitness data, from daily steps and sleep cycles to heart rate and GPS-based metrics.

The design and placement of these sensors are thoughtfully tailored to cater to the specific demands of each sport or activity. Swimmers have sensors fitted to their goggles \cite{eisenhardt2023augmented}, while boxers use gloves embedded with sensors for detailed punch analysis \cite{cizmic2023smart}. For activities where precision in heart rate monitoring is crucial, sensors are incorporated into headbands \cite{moov}. For activities like yoga or gym exercises, sensors embedded into shirts and pants assess muscle engagement and provide haptic feedback to ensure proper posture \cite{hexoskin, wearablex}.

\subsection{Ubiquity and Convenience of Wearable Technology}

As wearable technology progresses, there's a growing emphasis on balancing innovative features with user convenience. Modern sensor advancements now allow commonplace devices to perform diverse functionalities even if the location, like the wrist, isn't always optimal. The pedometer, once requiring specialized equipment on the waist, now operates effectively from the wrist \cite{bassett2017step}. Sleep monitoring, initially requiring sophisticated sensors, is now a common feature on many commercial smartwatches \cite{stone2020evaluations}. Other functionalities like monitoring exercise, calculating calories burned \cite{us11638556b2}, and tracking hand wash frequency \cite{us20210063434a1} are seamlessly implemented to operate well even from the passive wrist. Such shifts reflect a trend where user comfort and ubiquity are paramount.

Numerous studies have explored the use of commercially available devices for diverse tasks, due to their ubiquity and user familiarity. IMUPoser employs IMU sensors in the iPhone, Apple Watch, and AirPods to investigate body pose estimation, a challenge typically addressed with professional sensors \cite{mollyn2023imuposer}. Similarly, Pose-on-the-Go utilizes the camera and embedded sensors of the iPhone for the same purpose \cite{ahuja2021pose}. Shen et al. highlight the potential of a single smartwatch to capture the 3D pose of the arm \cite{shen2016smartwatch}, while DeVrio et al. investigate the use of UWB and IMU data available from a smartphone and a smartwatch for the same purpose \cite{devrio2023smartposer}. Further, research on using smartwatches to detect and recognize sports activities spans sports like table tennis \cite{vu2018smartwatch}, swimming and badminton \cite{zhuang2019sport}, and tennis \cite{taghavi2019tennis, lopez2019site}. Additionally, numerous studies highlight that smartwatches can distinguish between everyday activities such as opening doors and eating \cite{tchuente2020classification, laput2019sensing}.

Efforts to integrate wearable technology more seamlessly into their usage patterns are evident in recent studies. For instance, IMUPoser not only adjusts to various device placement locations but also accommodates differing numbers and combinations of devices in use \cite{mollyn2023imuposer}. In the realm of smartwatches, the placement --- whether on the dominant or passive arm --- has been explored. Mirtchouk et al. explored the use of body-worn sensors to recognize eating actions, finding that motion data from either wrist performed well when accompanied by audio data from earbuds \cite{mirtchouk2017recognizing}. Tchuente demonstrated high performance in classifying aggressive and non-aggressive activities for both wrists, with the dominant arm achieving slightly better results \cite{tchuente2020classification}. Similarly, Cvetovic found that the dominant arm's wrist allowed for a more nuanced classification of actions compared to the passive arm's wrist \cite{cvetkovic2017recognizing}. Haider et al. compared both wrist locations for volleyball action modeling, indicating slightly higher accuracy when using data from the passive arm, possibly due to the dominant arm's involvement in auxiliary motions \cite{haider2019evaluation}. Dieu et al. showed no significant differences in the measured physical activity from each wrist, challenging the assumption that the dominant arm, despite being stronger, significantly influences activity detection \cite{dieu2017physical}. These studies collectively emphasize the importance of aligning wearable technology with users' preferences and behaviors.
\section{Preliminary Survey}

To understand the perceptions and practices of wrist-worn device use among recreational tennis players, we conducted a preliminary survey on watch/smartwatch ownership. Our cohort comprised 40 players (12 female, 28 male, mean age 25.6) with over a year of tennis experience, gathered from our institution.

Among the respondents, one participant was left-handed, and the rest were right-handed. Out of the 40, 25 individuals reported owning and regularly wearing either a smartwatch or a regular watch (smartwatch: 22, regular watch: 3), all of which were worn on their passive (non-dominant) arm. Of those who wore watches, 84\% affirmed wearing them during tennis play for reasons such as tracking physical activity (e.g., heart rate, active calories burned), staying accessible for urgent calls, ensuring constant time-checking, and perceiving it as a non-burdensome accessory.

Additionally, we presented the assumption that a smartwatch could analyze their tennis shots and inquired about their willingness to use such a device, along with their preferred wear location. Impressively, 93\% expressed willingness to wear a smartwatch during play for enhanced functionalities. The primary motivations were to gain insights for skill improvement, self-evaluation, and capturing detailed performance information beyond what video recordings could provide.

Regarding the preferred location of the device, 70\% favored the passive arm, emphasizing comfort and ease. The remaining participants who preferred the dominant arm centered around the belief that it would allow a more accurate capture of their motions. Nonetheless, all respondents who wore watches during play consistently placed them on the passive arm, highlighting the prevailing preference for comfort and minimal interference with their natural motion during tennis.
\section{\sysname{}}

We propose \sysname{}, which aims to track tennis shots from the passive arm so that it is unobtrusive to the play while still giving shot analysis. We focused on six fundamental tennis shots: \textit{smash}, \textit{serve}, \textit{forehand stroke}, \textit{backhand stroke}, \textit{forehand volley}, and \textit{backhand volley}, representing the essential shot categories in tennis. Although shots can be subdivided by spin variations such as topspin or backspin, we generalized the categories to accommodate players of varying skill levels. Additionally, both one-hand and two-hand backhand strokes were categorized as \textit{backhand stroke}, despite their different passive arm movements, because these strokes generally fall under the same stroke category.

\sysname{} focuses on the fundamental analysis of tennis shots: (1) Shot Detection and (2) Shot Classification. Based on the passive arm data, Shot Detection first identifies all instances of a shot within a continuous tennis rally. Shot Classification then determines the shot type among the six shots from the captured segment of data. The analysis is then summarized and shown to the user through a prototype application.

To achieve this, we first collected motion data using IMU sensors from 20 recreational tennis players. Then, we developed a neural network model for shot detection and classification. Finally, we created an end-to-end prototype system utilizing a commercial smartwatch that shows the analysis results through a mobile application. Below we explain each component in detail.

\subsection{Dataset Collection}
\begin{figure}[b]
    \centering
    \includegraphics[width=\linewidth]{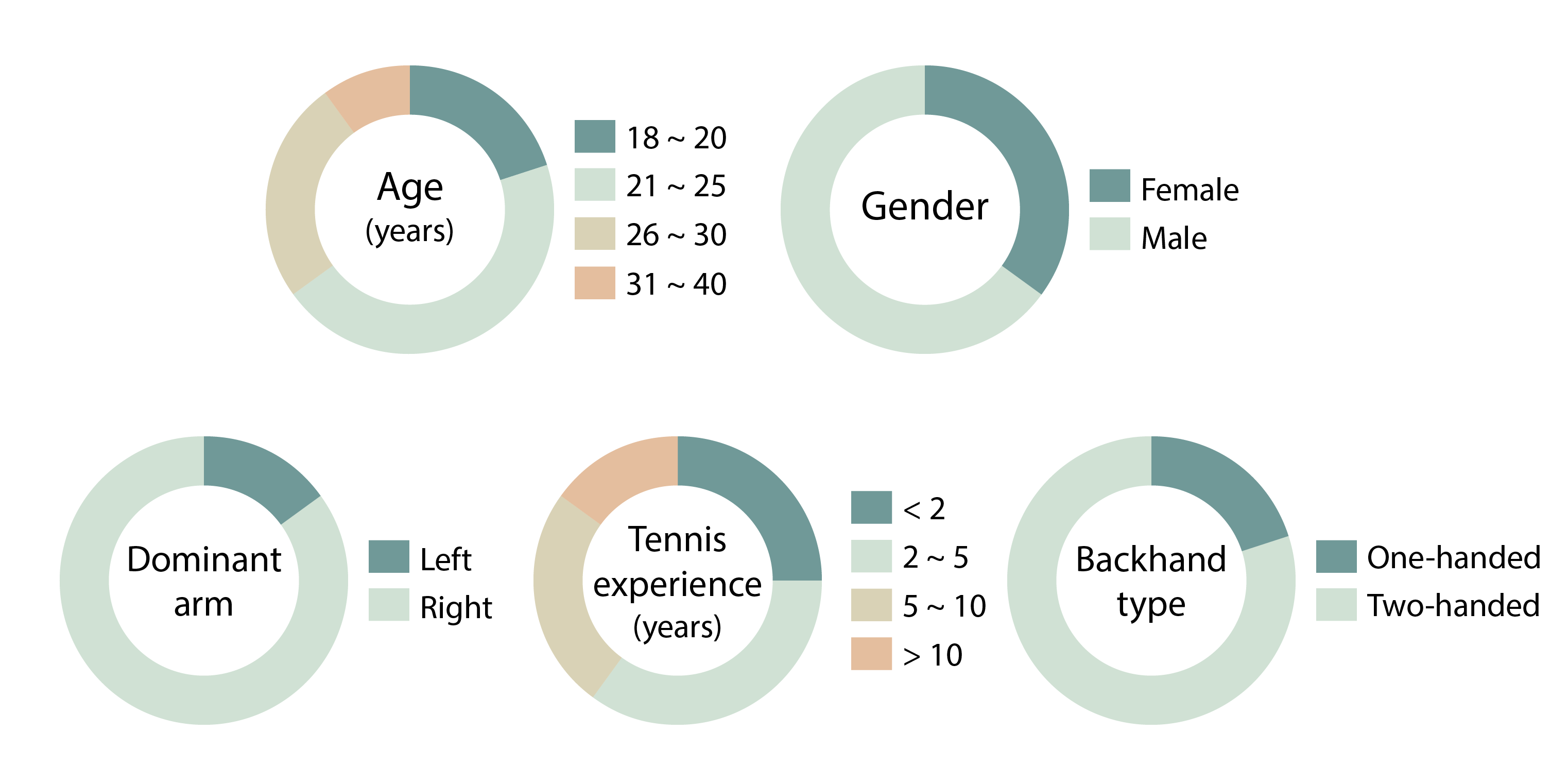}
    \caption{Participant demographics and tennis characteristics}
    \label{fig:participant_demographic}
    \Description{}
\end{figure}

We recruited 20 recreational tennis players representing diverse ages, genders, and skill levels, from tennis clubs within our institution (Figure ~\ref{fig:participant_demographic}). A minimum of six months of tennis experience was required to ensure that participants could execute the shots required in this study.

\begin{figure}[t]
    \centering
    \includegraphics[width=.8\linewidth]{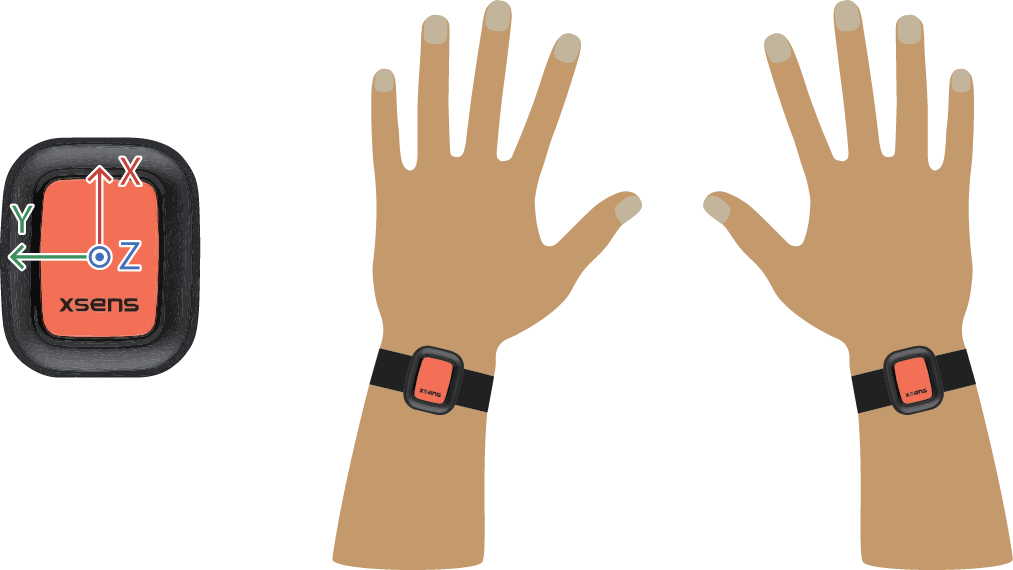}
    \caption{Sensor placement and orientation. For both the dominant and the passive arm, the IMU is placed with the positive X-axis pointing out towards the hand.}
    \label{fig:sensor}
    \Description{}
\end{figure}

We collected motion data using IMU sensors attached to both the dominant and passive wrists. Specifically, we used the Xsens DOT sensor \cite{xsens} to collect the motion data. The sensor measured 3-axis linear acceleration ($m/s^2$) and 3-axis angular velocity ($degree/s$). Figure ~\ref{fig:sensor} shows the orientation of the sensor on each wrist. To align the data of left-handed participants with that of right-handed ones, the Y-axis acceleration values and the X- and Z-axis angular velocity values were inverted by multiplying them by -1. The two sensors (on the dominant and passive arms) were synchronized and set to a frequency of 120 Hz. 

The dominant arm’s data was utilized to automatically detect shot instances by setting a simple threshold similar to Ganser et al.~\cite{ganser2021classification}. The sum of the squared values of each axis' acceleration was calculated to represent the relative force exerted at each timeframe. This produced a peak at each moment of impact due to the abrupt change caused by the impact of the ball. The time of the highest peak point was then assigned as the impact point. At each impact point, we extracted a shot window of 1.5 seconds (180 frames), composed of 1 second before the impact and 0.5 seconds after. Simultaneously, a camera was set up to record the sessions, serving two primary purposes: labeling the shot types and filtering out shots incorrectly detected by the threshold (e.g., when picking up a ball or hitting their own feet with the racket).

\subsubsection{Shot Classification}
For the shot classification dataset, 20 participants performed 50$\sim$60 shots for each of the six tennis shot types, following their preferred sequence. For all shot types except serves, participants performed the shots in a ball-feeding setting. In this setup, another person hit a ball towards the participants from the other side of the net, for them to hit back using the target shot type. A ball was fed roughly once every four seconds and participants were allowed to take a break whenever needed. For forehand and backhand strokes, participants stood near the baseline, while for forehand and backhand volleys, they stood approximately one meter from the net. Smash shots were executed with participants positioned near the service line, and for serves, participants, equipped with a box of balls at the baseline, were instructed to hit serves from either side of the court. This setup ensured a balanced dataset by controlling the frequency of each shot type, thus avoiding biases toward certain shot types (e.g., strokes more than volleys) often observed during rallies.

After data collection, the lead author watched the recorded video in conjunction with the sensor data to label each automatically detected shot and select the first 50 successful shots of each type. This resulted in a final dataset consisting of 6000 (=50*6*20) shot sequences. Each shot is labeled with the shot type and the hitter’s participant ID, and the mapping between the ID and their information (age, gender, tennis experience, dominant arm, backhand type) is recorded separately.

\subsubsection{Shot Detection}
The shot detection dataset contains long continuous sequences within which various tennis shots are hit at arbitrary intervals. Data was collected from participants playing rallies or casual matches. Contrary to the shot classification dataset where the same shot is hit at regular intervals, rallies and matches contain different types of shots being hit at different intervals, with more dynamic movements due to the ball coming at various speeds and positions. In addition, these sequences encompass incidental racket actions like hitting a ball off the ground or stopping a ball from flying out, as well as periods of inactivity where the racket is not used, such as when players pick up balls or discuss the score. Throughout this session, participants were not bound by predetermined sequences; instead, they were free to engage in rallies and matches according to their own preferences.

After data collection, the lead author watched the recorded video in conjunction with the corresponding sensor data sequence to remove all incorrectly detected shot windows. Then, all frames that were within a remaining correct shot window were labeled as `1' while the rest were labeled as `0'. Of the 20 participants, 10 participated in this dataset, resulting in a final dataset of 368 minutes of rallies consisting of 2259 shots.

\begin{figure*}[t]
    \centering
    \includegraphics[width=0.9\linewidth]{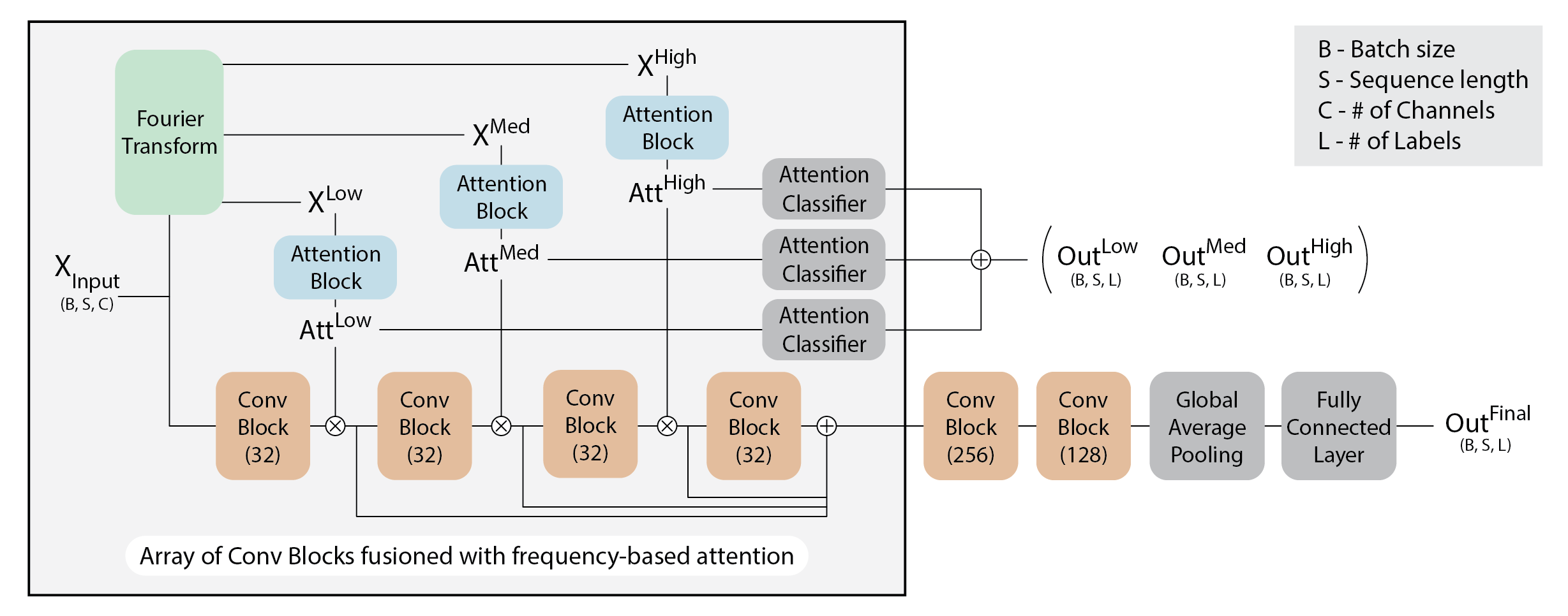}
    \caption{Model architecture for shot classification. To capture maximal information from the motion of the passive arm, the Fourier Transform is used to decompose the signal into low, medium, and high frequencies. An attention feature is generated from each frequency subpart which is then multiplied into the outputs of the convolution blocks one at a time. Shot class predictions are also generated from each attention feature. The Conv Blocks colored in orange represent a combination of a 1D convolutional layer followed by batch normalization and Mish activation function. The number within the brackets indicates the size of the output channel.}
    \label{fig:classification_model}
    \Description{}
\end{figure*}

\subsection{Neural Network Model}
We explored various neural network models but prioritized simpler architectures for shot classification and shot detection due to their computational efficiency. Although deeper architectures like transformers occasionally outperformed others, their advantages were marginal, and they demanded extended training duration.

 \subsubsection{Shot Classification Model}
The shot classification task determines what shot type the given IMU data represents among the six possible shot types. Drawing inspiration from prior work that leverages neural networks for classifying tennis shots from dominant arm data \cite{ganser2021classification}, our model consists of three 1-dimensional convolutional blocks. Each block incorporates batch normalization and employs the Mish activation function \cite{misra2019mish}. These blocks use a kernel size of 11 and sequentially expand the channel size to 128, then 256, and finally reduce it back to 128. A Global Average Pooling (GAP) layer follows to compact the feature maps, which are then passed to a fully connected layer with a softmax operation, generating a probability distribution across the six classes.

We introduced tailored modifications to the model to suit our focus on the passive arm. The model employs the Fourier transform to decompose the input signal into three frequency bands: the low-frequency band, targeting larger movements of the whole body; the medium-frequency band, focusing on motions of the upper body such as the arm's swing; and the high-frequency band, intended to capture any vibrations caused by the impact of the ball. These specific frequency ranges for each band (low - 0 to 4 Hz, medium - 5 to 20 Hz, high - above 20 Hz) were selected based on insights from previous studies in human motion analysis \cite{ji2005frequency, khusainov2013real}.

We refined the first convolutional block, partitioning it into four separate blocks with reduced channel sizes. Each frequency band from the Fourier transform passes through an \add{Attention Block --- a 1D convolution layer of kernel size 11 followed by a Sigmoid activation function ---} to generate a temporal attention vector with a channel size of 1. These vectors are then multiplied by the outputs from their corresponding previous convolutional blocks. Additionally, to ensure attention features capture relevant information for shot types, each attention feature is passed through an Attention Classifier block to produce class predictions, which are used for calculating the loss. The Attention Classifier block comprised a convolutional layer to re-expand the channel to 16, a ReLU activation function, a fully connected layer, and a Softmax activation function. Cross-entropy loss was used for all outputs. Figure \ref{fig:classification_model} illustrates the architecture of our modified model, with the adaptations shown inside the bordered left-side box.

\subsubsection{Shot Detection Model}

In tennis rallies and matches, the actual time spent executing a shot is relatively brief compared to the entire session, which also encompasses other movements and pauses between points. Thus, applying shot classification in real-time, or to all segments of data in a sliding window approach would result in an unnecessarily large number of inferences. Therefore, we design a model for shot detection that can compute over the entire sequence of data to identify the moments where a shot seems to have occurred.

While data from the dominant arm facilitates shot detection via straightforward threshold techniques --- owing to the pronounced jerk that occurs during racket-ball impact \cite{ganser2021classification} --- this is not easily replicated using data from the passive arm. Additionally, traditional threshold methods may fall short when detecting subtler shots, such as volleys or drop shots, where the racket-ball collision is comparatively gentle.

To address these challenges, we opted for the MS-TCN model \cite{farha2019ms}, a widely used convolutional layer-based neural network for action segmentation. This model excels at capturing temporal dependencies in sequences, allowing for more accurate frame-by-frame classification. We modify this model such that each frame within a given sequence is classified as either `true' (indicating it is part of a shot) or `false' (indicating it is not part of a shot). Our implementation consists of three stages, each with four layers, and utilizes a hidden dimension size of 64. We conducted training using cross-entropy loss and adjusted the loss function with a 5:1 class weight to account for the relative scarcity of shot instances in comparison to non-shot frames.

The frame-by-frame classification approach of our shot detection algorithm inherently leads to over-segmentation, where segmentation results may be noisy as neighboring frames are classified differently \cite{park2022maximization}. To refine the initial frame-wise detection results and extract more precise shot segments, we incorporate a refinement heuristic. Whenever a sequence of $k$ consecutive frames is classified as `true', a window of 180 frames is created centered at the midpoint of these consecutive frames. This process is first applied throughout the entire sequence. Then to avoid multiple extractions of a single shot, overlapping windows are merged into a single window that is centered at the mean of the overlapping windows' centers. This ensures that all shot instances are of equal length, facilitating shot classification within these extracted windows.

\subsection{Prototype Application}
\sysname{} is an end-to-end prototype system utilizing a commercial smartwatch and cloud server to demonstrate the design opportunities that our work enables. The system records IMU sensor data as the user engages in a tennis rally or match and subsequently tracks each executed shot. Afterwards, the user is provided with a timeline of the shots executed during the session along with the frequency of each shot type executed.

\begin{figure*}[t]
    \centering
    \includegraphics[width=0.9\linewidth]{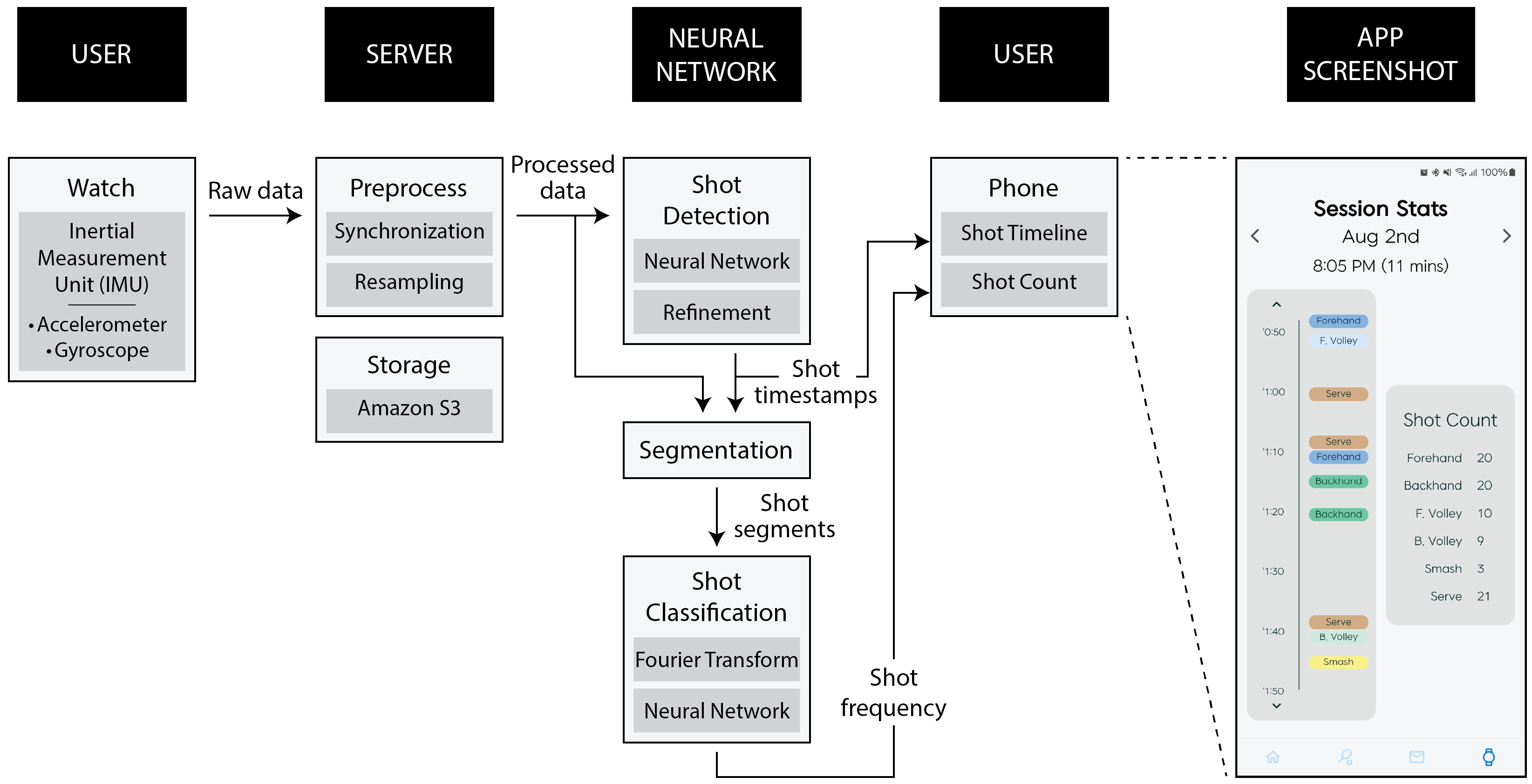}
    \caption{System architecture of \sysname{}. When the user initiates the app on their watch, the accelerometer and gyroscope readings from the IMU are recorded and sent to the server, which is then preprocessed to be inputted to the neural network. The Shot Detection module first detects all instances of shots within the given data, returning the timestamps of all shots. The Shot Classification module then uses the list of shot segments to generate a class prediction for each shot. The shot timestamp and shot frequency results are then sent to the user's phone, for it to display a timeline of all the shots the user performed during the session, as well as an overview of the number of shots for each type.}
    \label{fig:system_architecture}
    \Description{}
\end{figure*}

\subsubsection{System Overview}
Figure \ref{fig:shot_tracker_flow} illustrates the general user flow of \sysname{}. Before users start a tennis rally or match, they simply press the `Start' button on the app interface on the smartwatch worn on the passive arm. This prompts the smartwatch to start recording its accelerometer and gyroscope data. The collected sensor readings are then transmitted to the Amazon Simple Storage Service (S3) \cite{awsS3} in real-time. Upon the session's completion, users can end data collection by pressing the `Stop' button on the smartwatch interface. This triggers the server to process the accumulated sensor data to detect and classify all shots that were executed during the session. The results can then be viewed on an accompanying smartphone application (Figure \ref{fig:system_architecture}).

For each rally or match session that the user records, a breakdown of the shots they hit during the session is displayed, along with the date, time, and duration of the session. Using the results of Shot Detection, a timeline of all the shots that the user hits is plotted, giving an idea of the pace of the session. Each of the plotted shots is labeled by the type of shot it was, obtained through the results of Shot Classification, providing users with an overview of how their shot types changed as the rally or match progressed. Next to the timeline, a general tally of each shot type for the whole session is provided for an easy overview.

\subsubsection{Implementation}
We developed the application for the Samsung Galaxy Watch 4 using the Android Kotlin framework. The smartwatch's \texttt{TYPE\textunderscore ACCELEROMETER} and \texttt{TYPE\textunderscore GYROSCOPE} sensors were set to capture data at their maximum frequency of 100Hz. To ensure compatibility with our training data, we up-sampled the sensor readings to a 120Hz frequency through linear interpolation and adjusted the axes to align with the coordinate system of the Xsens DOT.
\section{Pipeline Evaluation}
We begin by evaluating the performance of our pipeline designed for shot classification and detection and compare the performance depending on the passive and dominant data. Then, we compare the effects various factors have on the performance to better understand the potential of using data from the passive arm.

\subsection{Experiment Setup}
All data was normalized to a range between 0 and 1 for training. Separate scalers were applied for linear acceleration and angular velocity. The shot classification model underwent 100 epochs of training with a batch size of 64 and a learning rate of 1e-4. The shot detection model was trained for 500 epochs with a batch size of 1 and a learning rate of 1e-3. Both models were trained using the Adam optimizer. We used the PyTorch framework and trained both models on an NVIDIA GeForce RTX 3090 GPU. \add{Additional details on model size and computation time are provided in the Appendix (Section~\ref{appendix_model})}

To accommodate the high degree of inter-subject variability, we adopted a 5-fold cross-validation approach, where three folds are used for training, one for validation, and one for testing. For the shot classification task, the 20 participants were randomly divided into five groups of four, while ensuring a balanced representation of tennis experience, gender, backhand type, and dominant arm across the folds. For the shot detection task, the 10 participants were randomly divided into five groups of two. For both tasks, the performance of each fold was averaged out to obtain the final performance. For shot classification, we used accuracy as the evaluation metric, while for shot detection, we incorporated the F1 score for positive labels in addition to the frame-wise accuracy, to address the unbalanced ratio of frames containing shots versus those without.

\subsection{Results}

\subsubsection{Shot Classification}
Table \ref{tab:classification_table} presents the results of the shot classification task. We compare our model with the original FCN model by Ganser et al. \cite{ganser2021classification} which we utilized as our backbone. Our shot classification model achieved an average accuracy of 88.2 ± 2.0\% with the passive arm data and 90.1 ± 3.0\% with the dominant arm data. As anticipated, the dominant arm data exhibited slightly superior performance, with a marginal difference of 1.9\%. In contrast, the FCN model displayed a notable performance gap between the passive and dominant arm data, amounting to 9.1\%, emphasizing the greater difficulty in distinguishing shots from the passive arm. The introduction of frequency-band attention modules in our model mitigated this gap, improving performance by 6.8\%. Interestingly, this enhancement had no significant impact when using the dominant arm data, suggesting that the attention modules compensated for the absence of information from the passive arm.

\begin{table}[t]
\begin{center}
\begin{tabular}{ l c c c}
\hline
 & \textbf{Passive} & \textbf{Dominant} & \textbf{\textit{Diff}} \\ 
\hline
 \textbf{FCN \cite{ganser2021classification}} & 81.4 & 90.5 & \textit{9.1} \\
 \textbf{Ours} & 88.2 & 90.1 & \textit{1.9} \\
\hline
 \textbf{\textit{Improvement}} & \textit{+6.8} & \textit{-0.4} & \\
\hline
\end{tabular}
\end{center}
\caption{Comparison of the backbone model FCN by Ganser et al. \cite{ganser2021classification} with our modified model for shot classification.}
\label{tab:classification_table}
\end{table}

\begin{figure}[t]
    \centering
    \includegraphics[width=.85\linewidth]{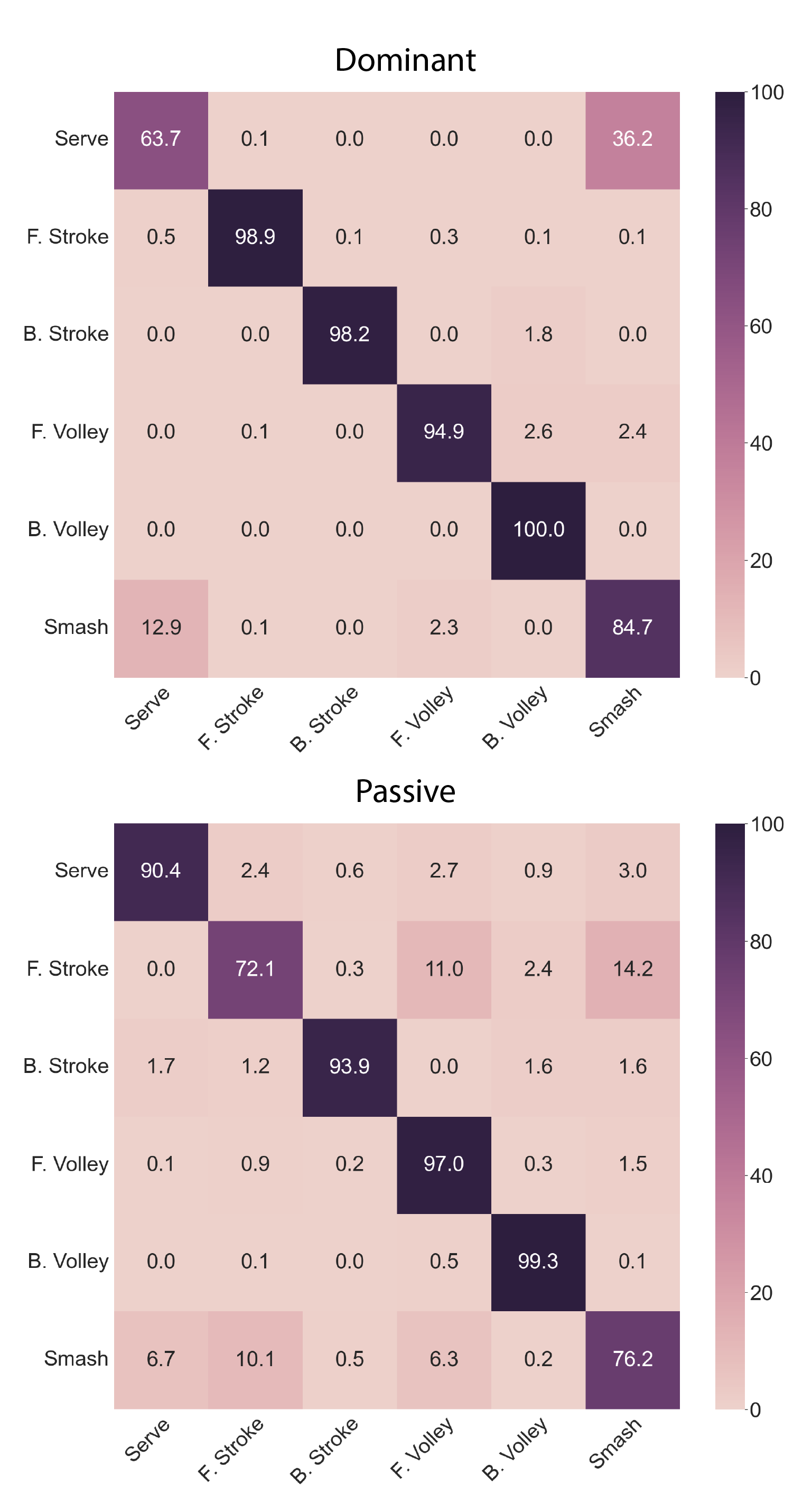}
    \caption{Comparison of the confusion matrices for shot classification using the dominant and passive arms' data. Using the dominant arm's data produces a higher classification accuracy for the "F.Stroke" class, while the passive arm's data results in a higher accuracy for the "Serve" class.}
    \label{fig:res_classification}
    \Description{}
\end{figure}

\begin{table}[b]
\begin{center}
\begin{tabular}{ l c c c c}
\hline
 & \multicolumn{2}{c}{\textbf{Passive}} & \multicolumn{2}{c}{\textbf{Dominant}} \\ 
 & Acc & F1 & Acc & F1 \\ 
\hline
 \textbf{Peak detection \cite{ganser2021classification}} & 77.8 & 37.6 & 97.2 & 90.1 \\
 \textbf{Ours} & 95.6 & 86.0 & 98.5 & 94.8 \\
\hline
\end{tabular}
\end{center}
\caption{Comparison of the threshold-based peak detection algorithm by Ganser et al. \cite{ganser2021classification} with our MS-TCN based model for shot detection.}
\label{tab:detection_table}
\end{table}

Figure \ref{fig:res_classification} presents the confusion matrix for both datasets across the six classes. The most significant difference is observed in the "forehand stroke" and "serve" classes. While the "forehand stroke" class is accurately distinguishable (98.9\%) from the dominant arm data due to the distinctive trajectory of the dominant arm, classifying the "forehand stroke" from the passive arm is considerably more challenging (72.1\%) due to potential variations in passive arm motion (see Figure \ref{fig:pro_form}). Surprisingly, confusion between the "smash" and "serve" classes was more prevalent in the dominant arm data, resulting in a lower average accuracy of 74.1\% compared to 82.8\% from the passive arm data. Both the "smash" and "serve" actions involve swinging the dominant arm over the head, resulting in similar motions, whereas the passive arm introduces a temporal distinguishing factor (e.g., tossing the ball with the passive arm right before a "serve"). These findings highlight that the challenge of distinguishing between similar motion patterns is not exclusive to the passive arm.

\begin{figure*}[t]
    \centering
    \includegraphics[width=0.95\linewidth]{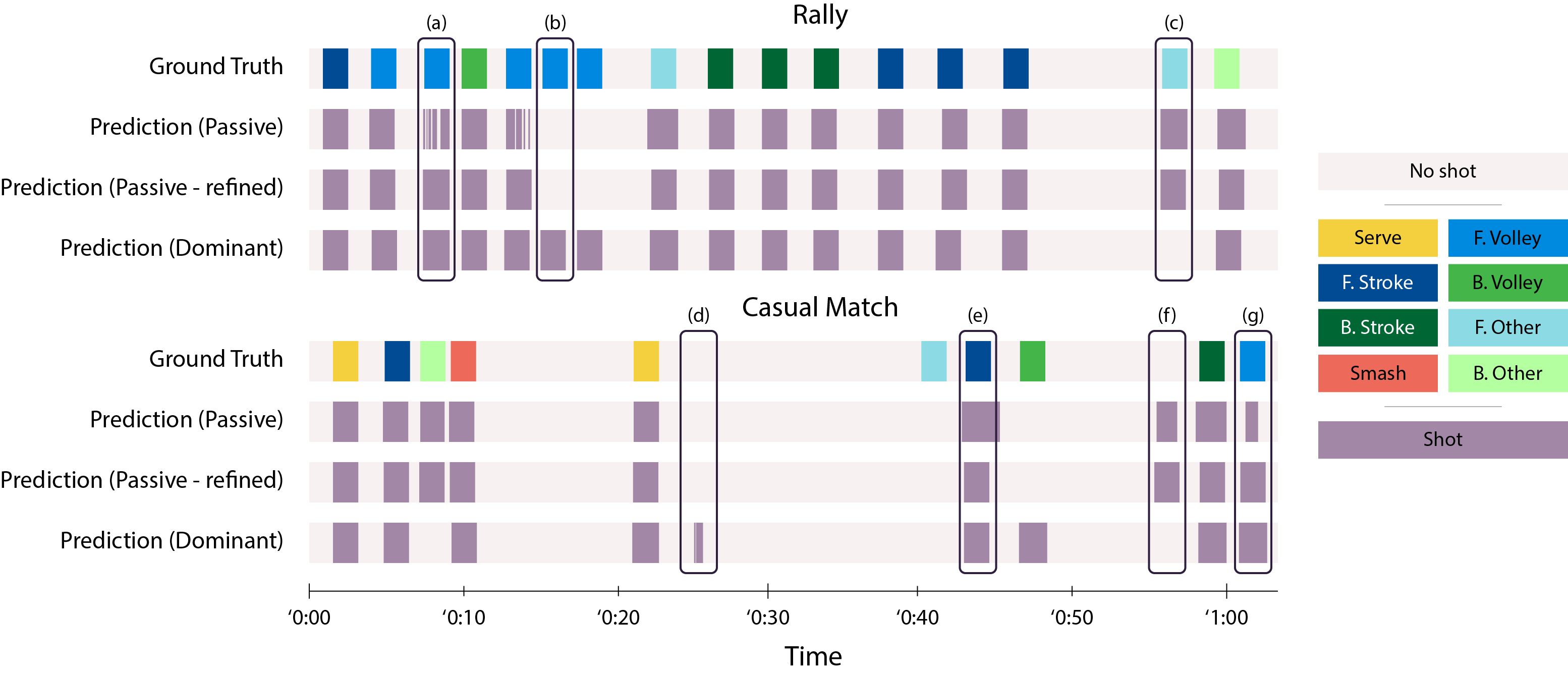}
    \caption{Shot detection results. Each timeline visualizes the predicted or ground truth labels of each frame within the sequence. The timings of correctly detected shots are generally precise. While both modes display some undetected shots (b \& c), the passive arm results in slightly more false negatives, especially for the "volley" classes. The detection refinement algorithm refines granular predictions into complete segments (a) and also adjusts the length of predicted segments to match the predefined length (e \& g). False positives are also present in both cases (d \& f).}
    \label{fig:res_detection}
    \Description{}
\end{figure*}

\subsubsection{Shot Detection}
Table \ref{tab:detection_table} displays the results of the shot detection task. We compare our approach with the threshold-based peak detection algorithm that is commonly used for detecting shots from the dominant arm \cite{ganser2021classification}. As expected, while the peak detection algorithm accurately detects shots using dominant arm data, it experiences a substantial drop in performance with passive arm data. This discrepancy arises because, aside from two-handed backhand strokes, the jerk created by the impact of the ball on the racket cannot be sensed from the passive arm. Our model outperforms the peak detection algorithm significantly, achieving an F1 score of 86.0\% from passive arm data. This underscores the potential of neural networks to capture distinguishing features between shots and non-shots from the motions of the passive arm. However, it still exhibits a notable gap of 8.8\% compared to the dominant arm, which attains an F1 score of 94.8\%.

Figure \ref{fig:res_detection} provides a visualization of the shot detection results during both rally and casual match scenarios. In rallies, the density of detected shots appears marginally higher compared to casual matches. Notably, when shots are successfully detected, the timing of the identified shot window is generally precise, despite some instances of slight misalignment with the labeled shot timings. The effects of the refinement heuristic are also illustrated. When the original predictions are granular, the refinement algorithm successfully fills up the frames to produce whole segments (a). It also corrects the length of longer or shorter segments that occur due to ambiguous shot timings, adjusting them to the predefined length (e \& g). However, it is important to note that the refinement algorithm's primary focus is to ensure shot segments of equal length, rather than to improve detection performance. The increase in F1 score introduced by the refinement algorithm was only around 0.2\% for both passive and dominant arm data, indicating that granular predictions or incorrect segment lengths are not the primary contributors to inaccurate detection.

False negatives, or undetected shots, are present in both results. In the case of the passive arm, these often manifest as volleys. This can be attributed to the smaller, more subtle movements made closer to the net, which can easily be mistaken for inactivity, particularly given the absence of noticeable jerks in the passive arm data. Conversely, the dominant arm data is also not immune to false negatives and positives, predominantly in the "Forehand Other" and "Backhand Other" classes. These tend to occur during actions such as picking up the ball, passing it to the opponent, or blocking it from going out with the racket. Because the dominant arm data-focused model is attuned to the jerk caused by ball impacts, these specific scenarios could be misinterpreted as "Other" shots which generally pertain to defensive strokes or drop shots.

\subsection{Ablation Study}

For a wider understanding of the use of the passive arm, we conducted an ablation study on the shot classification task. We investigated the effects the context in which the tennis game was played (either during ball feeding or in rally and match settings) had on the classification performance, as well as the length of the shot segment. \add{Additional experiments on the comparison of linear and angular data, the effects of fine-tuning to user-specific data, and variations in sampling frequency are reported in the Appendix (Section~\ref{appendix_experiments})}.

\subsubsection{Ball Feed vs Rallies and Matches}
Rallies and matches introduce more dynamic movements compared to ball-feeding sessions. To explore the influence of the playing context, we isolated and labeled all detected shots from rallies and matches. Shots that didn't fit into one of our six predefined categories --- such as drop shots, forehand slices, and lobs --- were excluded from this analysis, leaving us with a total of 1826 shots (Serve: 433, Smash: 20, F. stroke: 880, B. stroke: 344, F. volley: 82, B. volley: 67).

When the model, initially trained on ball feed data, was tested with the rally and match data, it achieved an accuracy of 86.0\%, 2.2\% lower than the average accuracy obtained from ball feed data. 
This discrepancy suggests a strong correlation between the motions captured during ball-feeding sessions and those captured in rallies and matches. Subtle differences seem to exist due to the additional dynamic movements inherent in rallies and matches, such as running or sliding while hitting the ball.

\subsubsection{Segment Length}
In this study, we utilized segments of 1.5 seconds (180 frames) that capture 1 second of motion before the ball impact to 0.5 second after. We investigated the influence segment length has on performance by comparing three alternate segment durations: 1-second segments (impact at the 0.5-second mark), 2-second segments (impact at the 1-second mark), and 2-second segments (impact at the 1.5-second mark).

As anticipated, employing 1-second segments resulted in a notably lower accuracy of 79.4\%, indicating that distinguishing features for each shot class emerge earlier in the shot motion. Utilizing 2-second segments yielded similar yet slightly reduced accuracies of 85.9\% and 87.3\% compared to our 1.5-second segments. This suggests that providing excessive data of the motion before or after each shot does not necessarily enhance performance and may rather confuse the model, as these motions may have weaker correlations with the type of shot executed. Therefore, selecting an appropriate segment length is crucial for accurate classification.
\section{User Evaluation}
We evaluate the practicality and efficacy of \sysname{} through a user evaluation study. The study aims to measure how feasible and practical \sysname{} is in a real-world setting and how efficient it is compared to existing methods.

\subsection{Study Design}
To evaluate \sysname{}, we used a within-subject design with two conditions: (1) playing tennis with our prototype smartwatch on the dominant arm, and (2) with the smartwatch on the passive arm (\sysname{}). For the Dominant arm condition, we used a neural network model of the same architecture, trained on the dominant arm's data. Everything else remained the same as the \sysname{} condition. 
The order of conditions was counterbalanced. For each condition, participants wore our prototype smartwatch on their arm while freely engaging in tennis rallies and casual matches for at least 5 minutes. After each session, we showed them our app showing the result of their tennis shot analysis, and they were asked to complete a survey about the mental and physical load (as part of the NASA-TLX questionnaire) and how it affected their tennis performance. Finally, we conducted a semi-structured interview asking about the overall experience and their preferences.

\subsection{Participants and Apparatus}
We recruited 10 participants (8 male, 2 female, mean age = 26.4) from our institution, all with at least a year of tennis-playing experience. On average, participants had 4.1 years of tennis experience (SD = 2.3). Among the 10 participants, 3 reported that they typically do not wear watches, while the other 7 indicated wearing a watch on their passive arm, with 6 wearing watches during tennis play as well. For the study, participants wore a Samsung Galaxy Watch 4 equipped with our IMU recording app. After each condition, the result was provided through our app on a Galaxy S8.

\subsection{Results}

\begin{figure}[t]
    \centering
    \includegraphics[width=\linewidth]{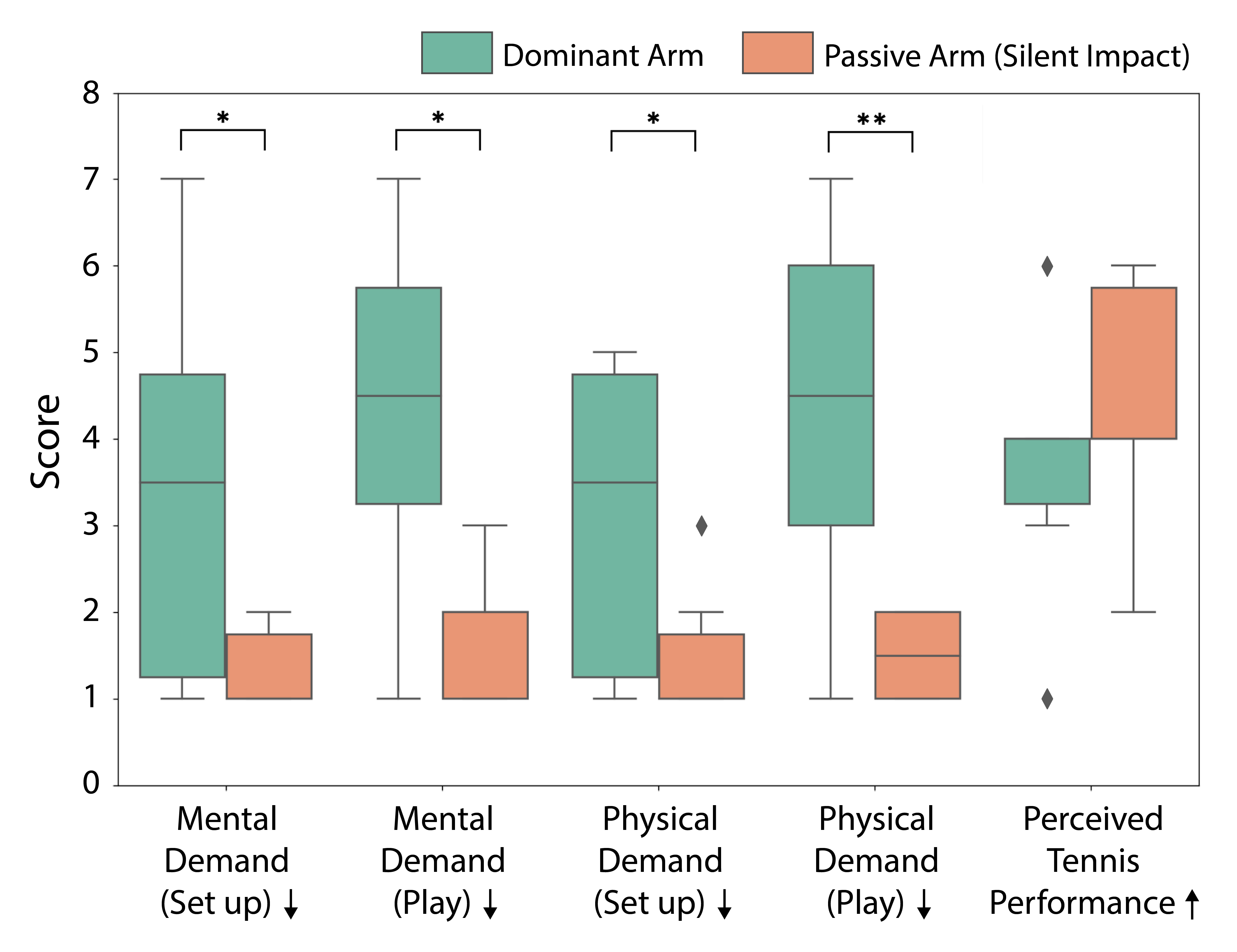}
    \caption{User rating on the mental and physical load as part of NASA TLX Load Index questionnaire, and perceived tennis performance. Participants felt less mental and physical demand with \sysname{} by a statistically significant level, in both the setup process and playing tennis. Also, they felt their tennis performance was better with \sysname{}, although the difference was not significant (*=p<0.05, **=p<0.01).}
    \label{fig:user_study_results}
    \Description{}
\end{figure}

Our results demonstrate that participants experienced significantly less mental and physical burden when using \sysname{} (Figure~\ref{fig:user_study_results}). This can be attributed to factors both during the device setup process and during play.

\subsubsection{Device Setup}
All participants appreciated the easy setup of the system. According to questionnaire responses rating mental and physical load on a scale of 1 to 7, participants felt less mental (Mann-Whitney U test, U=79, p=0.02) and physical (U=78, p=0.03) demands during the setup process, which includes wearing the smartwatch, in \sysname{} compared to the baseline. Those accustomed to wearing a watch while playing tennis noted that using \sysname{} felt no different except for possibly turning the app on. Participants who did not own smartwatches mentioned that if a wristband with the functionality were available, they would not be burdened putting it on, though some admitted they might forget occasionally. A few participants accustomed to wearing extra apparel like wrist sweatbands or support straps stated that it would feel similar to adding extra gear.

\subsubsection{During Play}
Participants reported that playing tennis with \sysname{} was less mentally (U=83, p=0.01) and physically (U=85, p=0.07) distracting. Those who regularly wore smartwatches while playing tennis experienced no physical discomfort or burden with our system, as it required no additional devices or changes to device placement. One participant accustomed to removing their watch during tennis initially felt a difference but quickly adapted to having the device on their passive arm. Those not accustomed to wearing smartwatches felt a more noticeable difference in the weight and balance of the arm. However, this only resulted in minor discomfort during backhand swings when the wrist needed to bend backward.

In contrast, wearing the device on the dominant arm was generally uncomfortable for most participants. Issues included the back of the hand touching the device during motions that require extensive wrist movement, such as forehand strokes or serves. Some participants also expressed discomfort due to the tightness of the watch strap, affecting wrist flexibility. The weight of the device was another concern, with one participant feeling that swings became heavier when the device was on the dominant arm. Conversely, participants accustomed to wearing wrist sweatbands on their dominant arm were less bothered by either the weight or the tightness of the strap.

Finally, although not statistically significant (U=31, p=0.12), participants felt their perceived tennis performance was lower in the baseline condition compared to when using \sysname{}. This suggests that \sysname{} had minimal impact on their tennis performance. Overall, our results demonstrate the practicality of \sysname{}, as it minimally interrupts play while providing valuable shot analysis.

\subsection{Feedback on Functionality}
Our prototype application records and presents shot counts and timelines for each shot's type and executed time. Feedback on these features varied, with some finding the basic statistics enjoyable for a quick game overview. Some mentioned that the summary of shot counts would aid in understanding the overall play style and balance of the match. One participant said, "I could see how aggressive my play was by looking at the number of volleys and smashes". Other participants appreciated the timeline feature, noting that it would help them remember how the match progressed.

Participants provided suggestions for enhancing the utility of the application, including additional details for each shot, such as whether it was a point-winning shot or an error, or a breakdown of shots into its power or spin. For personal improvement purposes, participants suggested tracking the swing trajectory and providing feedback for shot improvement. Despite these diverse opinions, participants uniformly appreciated the app's non-intrusiveness, likening it to other passive smartwatch features like calorie tracking. Most expressed willingness to let it run passively on their smartwatches during tennis play as it required no additional setup.

\section{Discussion}
Our study successfully demonstrated the feasibility of utilizing data from the passive arm for tennis shot analysis. We discuss how our approach enables various applications, how it can be generalized to other sports, and several opportunities for further improvements.

\subsection{Potential Applications of \sysname{}}

\sysname{} detects and classifies tennis shots from the passive arm. As briefly discussed in the previous section, participants from the user study felt that even this simple functionality can be valuable in various ways such as providing a summary of the gameplay or helping remember how the match progressed.

\sysname{} can also lay the groundwork for a wider range of useful and entertaining applications. For instance, engaging games could be designed, where players must execute specific shot types during a match or force opponents to exceed a designated shot frequency. These games would not only inject a fresh layer of enjoyment and challenge into tennis but also aid users in refining their game control. Moreover, analyzing the sequence of shots played within a match can offer valuable insights for those working on match strategies. Specific patterns of shots, such as a "serve and volley" strategy, can be identified from the shot sequence and the game's pace and speed can be determined from the density of shots on the timeline. All these functionalities are achievable simply by tracking which shots were hit and when.

Collaborative settings where multiple players in a match utilize the technology simultaneously can offer valuable insights. The collective data could be used to induce additional information beyond what individual use provides. For example, determining the final shot of each point could be inferred by analyzing whether the opponent executed a subsequent shot. Additionally, the direction (left or right) or style (short, long, or high) of the executed shot could be approximated based on the type of the opponent's return shot (forehand vs backhand, volleys vs stroke vs smash). Furthermore, it can serve as a mutual validation mechanism by checking for improbable scenarios such as simultaneous shots by multiple players or back-to-back shots from the same team, potentially enhancing the overall accuracy of shot detection and classification.

While previous attempts at similar applications in the tennis industry have not gained widespread success, \sysname{}'s unobtrusive and hassle-free approach, requiring minimal setup, could alleviate user burden and thus introduce new forms of interaction within the sport.

\subsection{Further Expansions}

\subsubsection{Passive Arm’s Potential for Tennis}
In terms of sensor placement, while we found that the passive arm offers advantages in comfort and usability, its capabilities in capturing more advanced metrics remain an open question. Complex variables like ball speed or spin rate, previously targeted by solutions using the dominant arm, may pose challenges for accurate measurement from the passive arm. This limitation arises because precise racket trajectory or ball force exertion cannot be directly assessed. \add{However, the context surrounding the shot, captured from the motion relayed from the whole body, can provide characteristic patterns, as shots can display varying motion before (i.e., racket takeback) or after the swing (i.e., follow through), in both magnitude (strokes have larger motions than volleys) and direction (backhand volleys rotate the body towards the passive arm).} Moreover, broader categories of ball speed or spin type could potentially be estimated based on captured vibrations or subsidiary motions from the passive arm. Furthermore, tennis encompasses diverse aspects beyond shot-related parameters that impact performance, including running speed, reaction time, tactics, stamina control, and consistency. Many of these aspects involve other body parts or the entire body and are not confined to the dominant arm. Therefore, it's crucial to explore how the passive arm can seamlessly provide insights into these dimensions as well.

\subsubsection{Integration of Other Modalities}
Integrating other modalities, especially sensors commonly found in smartwatches, offers a promising avenue for enhancing the system's functionality. Audio input could provide valuable context, with the sound of the ball being struck aiding in shot detection, and players' vocalizations (e.g., grunts, cheers) helping determine the outcome of shots or duration of each point. Photoplethysmography (PPG) data could provide comprehensive information regarding a player's physical state. While the IMU sensor alone faces challenges in precise location tracking due to noise accumulation \cite{byun2019walking}, obtaining rough estimates under specific conditions like the confined dimensions of a tennis court or limited serving locations, might be feasible. \add{Furthermore, our approach could be integrated with camera-based solutions such as Swing Vision \cite{swingvision}. Predictions from the wearable can correct erroneous predictions from the camera caused by occlusions (e.g., people, net), distance, or lighting conditions, creating a self-supervised system where the two modalities label anomalous data to train the other. Moreover, wearables can capture nuanced motions like split steps that are too subtle for cameras, to provide feedback on a player’s reaction time or footwork.}

\subsection{Generalizability and Scalability}
\subsubsection{Use of the IMU Sensor}
The use of the Inertial Measurement Unit (IMU) sensor facilitates the potential applicability of this approach across various devices and form factors. IMU sensors are widely adopted in modern technology, seamlessly integrated into a multitude of devices ranging from smartphones to wearables like smartwatches, fitness trackers, and wireless earphones. They are esteemed for their robustness, as they solely measure linear acceleration and angular velocity, often obviating the need for additional calibration or setup, unlike sensors such as Electromyography (EMG) or strain sensors. Consequently, we anticipate this system to be compatible with the majority of commonly encountered wrist-worn devices, albeit recognizing potential minor variations in performance depending on the quality of the embedded IMU sensor. Furthermore, this approach may extend to other device formats such as smart gloves or smart rings, provided they are proximal to the wrist of the passive arm and thus capable of capturing analogous motions.

\subsubsection{Scope of Motion Proxies}
\add{We demonstrate that using the motions of the passive arm as motion proxies is both technically effective and user-friendly for tracking tennis shots. This can be similarly applied to other sports with significant arm movements, though the effectiveness and implications will depend on the activity's nature. For sports like golf or badminton, where larger movements involve a wider range of the body, the motion of the passive arm could contain necessary information to distinguish finer actions. In sports with more constrained gestures, subsidiary motions may only capture actions at a coarse level, which can still open possibilities for creative utilization. For example, in billiards, vibrations from the passive wrist could be used as proxies for shot occurrences, enabling a smartwatch application that automatically keeps score. Furthermore, the concept of subsidiary motion proxies can extend to other areas of the body beyond the passive arm. In swimming, watches may hinder arm movement, and a waist strap that measures minute tilts of the waist as motion proxies could provide a more unobtrusive solution to swim stroke tracking instead. Leveraging areas like the ears, pockets, or eyes, commonly occupied by devices such as wireless earphones or smartphones, could also make the technology more accessible. As such, subsidiary motions captured from different parts of the body can enhance the experiential aspect of interactive technology in sports \cite{elvitigala2024grand}.}

\subsection{Limitations and Future Work}
A notable limitation of our study is the relatively small sample size. While our results showcase the potential of the passive arm, a larger and more diverse participant pool would enhance the statistical robustness and generalizability of our approach. Future research should consider including participants of a greater variety of ages and skill levels to capture a broader spectrum of playing styles and preferences.

Furthermore, we focused on only six types of tennis shots for our data collection. This limitation restricts the depth of analysis possible. A more detailed categorization of shot types, including factors such as the spin, power, or direction of the ball (e.g., top-spin, slice, lobs, drop shots, cross-court shots), could provide players with a richer understanding of their performance and areas for improvement. Moreover, exploring the potential of the passive arm for more comprehensive analyses regarding both shots (e.g., shot powers, swing trajectory) and the broader movements of the body (e.g., location on the court, movement speed), could open up new dimensions in enhancing player experience.

We demonstrated the efficacy of neural networks in accurately detecting and classifying shots using passive arm data. However, this process necessitated the use of a powerful GPU for computation, leading our pipeline to transmit data to a server. Future investigations could explore the utilization of lighter neural networks or traditional machine learning algorithms, such as support vector machines while preserving high accuracy. This exploration could eliminate the reliance on a server, enabling all computations to occur directly on the user's device.
\section{Conclusion}

In this study, we challenged conventional methods of tennis shot analysis by introducing a novel approach that utilizes the passive arm's motion, demonstrating its feasibility for accurate shot classification and detection. Our findings reveal that this method not only is comparable in performance to traditional dominant arm-based solutions but also significantly enhances user comfort and mitigates the hindrance in movement often associated with wearables on the dominant arm. By developing a prototype application on a commercial smartwatch, we showcase a practical, unobtrusive, and user-friendly solution that opens new horizons for recreational tennis players aiming to analyze their performance effortlessly. This work sets the stage for future wearable technologies that can bring sophisticated sports analytics into everyday convenience.

\begin{acks}

This work was supported in part by the Ministry of Trade, Industry \& Energy (MOTIE, Korea) through Technology Innovation Program under Grant 20007058, and in part by the National Research Foundation of Korea (NRF) grant funded by the Korean Government (MSIT) under Grant RS-2023-00208052.

\end{acks}
\bibliographystyle{ACM-Reference-Format}
\bibliography{references}

\appendix
\clearpage

\section{Appendix}

\subsection{Additional Experiments}
\label{appendix_experiments}

\add{We report the results of supplementary experiments conducted to examine the effects of the type of inertial data, fine-tuning, and sampling frequency. Furthermore, we provide some technical details of our neural network model.}

\subsubsection{Linear Acceleration vs Angular Velocity}
IMU sensors contain both an accelerometer for measuring linear acceleration and a gyroscope for measuring angular velocity. We assessed the performance of each module individually and found that relying solely on one led to lower accuracy. Using only the gyroscope yielded an accuracy of 77.2\%, while the accelerometer achieved a slightly higher accuracy of 81.8\%.

Interestingly, the effectiveness of each module varied across classes. The accelerometer demonstrated significantly higher accuracy for the "forehand stroke" class (72.2\% vs. 51.6\%), suggesting that during forehand strokes, the rotation of the passive arm wrist varies considerably, whereas linear translation remains relatively consistent across individuals. Conversely, both "volley" classes exhibited higher performance with gyroscope data (average 94.4\% vs. 87.2\%). This finding aligns with expectations, as volleys involve less swinging of the passive arm and more rotation of the entire body.

\subsubsection{Fine-tuning with User Data}
\add{Since the pose can vary greatly between individuals, we investigated the effect of fine-tuning with the target user’s data. A model initially trained without the target user's data was then fine-tuned at a lower learning rate of 1e-6 using 10\% of the user’s data (5 instances of each shot). This led to an increase in accuracy by 4~7\% for each individual, with the overall average at approximately 94\%. This can benefit Silent Impact as users can provide a few instances of each type of shot to train a user-specific model that better captures their characteristic movements.}

\subsubsection{Sampling Frequency}
\add{To examine the robustness of our approach to lower-quality data, we trained and evaluated our model with data at lower sampling frequencies. The collected data was downsampled to 30 Hz and 60 Hz. In both cases, the changes in performance were negligible, with a reduction of less than 0.5\%.}

\subsection{Technical Details}
\label{appendix_model}

\subsubsection{Model Size}
\add{The shot classification model consists of 340K parameters, with 287K parameters belonging to the backbone model and 53K parameters to the attention modules. To determine if the improved performance of our model was solely due to the increased model size, we tested a backbone with an increased channel size to match the parameter count of our model. This resulted in only a 0.4\% increase in accuracy, showing that the performance improvement is primarily due to the attention modules}

\subsubsection{Computation Time}
\add{Experiments were conducted on a server equipped with an Intel Xeon 4310 CPU @ 2.10 GHz and an NVIDIA GeForce RTX 3090 GPU. For shot classification, our model took approximately 0.2 seconds to infer 100 instances of a shot, including performing the Fourier decomposition. For shot detection, our model took 0.3 seconds to infer on 10 minutes of data. However, these times will vary depending on the computational environment.}

\end{document}